\documentstyle[aps,epsfig,eqsecnum]{revtex}
\begin{document}
\draft

\title{Constraints, Histones, and the 30 Nanometer Spiral}

\author{Roya Zandi and Joseph Rudnick}

\address{Department
of Physics, UCLA, Box 951547, Los Angeles, CA 90095-1547}
\date{\today}

\maketitle

\begin{abstract}

We investigate the mechanical stability of a segment of DNA wrapped 
around a histone in the nucleosome configuration.  The assumption 
underlying this investigation is that the proper model for this 
packaging arrangement is that of an elastic rod that is free to twist 
and that writhes subject to mechanical constraints.  We find that the 
number of constraints required to stabilize the nuclesome 
configuration is determined by the length of the segment, the number 
of times the DNA wraps around the histone spool, and the specific 
constraints utilized.  While it can be shown that four constraints 
suffice, in principle, to insure stability of the nucleosome, a proper 
choice must be made to guarantee the effectiveness of this minimal 
number.  The optimal choice of constraints appears to bear a relation 
to the existence of a spiral ridge on the surface of the histone 
octamer.  The particular configuration that we investigate is related 
to the 30 nanometer spiral, a higher-order organization of DNA in 
chromatin.
\end{abstract}

\pacs{87.15.By, 87.15.La, 87.15.-v, 62.20.Dc}

\section{Introduction}
The issue of DNA packaging has been the subject of intense research
for the past forty years.  The remarkable fact that a meter of DNA
(the total length of the human genome) fits into a cell nucleus having
a typical radius of a few microns, and that, thus packed, still
manages to perform all its biological functions, has captured the
attention and continues to challenge the ingenuity of researchers.
The fundamental unit of DNA packing in eukaryotes is the nucleosome
\cite{korn,oudet}.  In the nucleosome configuration, a portion of the
DNA strand wraps approximately one and three quarter times around a
protein spool, known as a histone \cite{luger,klug1,rich}.  A string of
nucleosomes is believed to participate in the next higher order of DNA
packing by folding to form the so-called 30-nm fiber \cite{chrom}.
Even higher orders of organization have been conjectured, but as yet
there is nothing approaching a complete understanding of the physical
structure, at all orders, of the DNA in the cell nucleus.  Indeed, the
detailed oragnization of DNA and proteins in the 30 nm fiber is not
entirely settled \cite{bed,skep1,skep2,however}.

In this paper, our ultimate focus will be on a single nucleosome.  We
will present a model for the action of histones in nucleosomes.  We
treat the segment of DNA in a nucleosome as an elastic rod and apply
an approach first developed by Kirchhoff \cite{kirc} to obtain the
equilibrium configurations of DNA in the absence of histones.  We
assess the stability of the elastic rod with respect to small
deviations from the equilibrium configurations, and we find that all
configurations that are not equivalent to a straight rod are unstable
to fluctuations.  More specifically, all non-trivial equilibrium
configurations represent saddle-points in energy space.  Then we prove
that the presence of histones provides a physical mechanism by which
more compact configurations of DNA are rendered stable against purely
mechanical fluctuations.  The specific mechanism is a set of
constraints on the fluctuations about mechanical equilibrium, which
can be simply modeled mathematically.  The stability of the equilibrium
configuration is then framed in terms of the determinant of an
$n$-by-$n$ matrix, where $n$ is the number of constraints.  Of special
interest to us is the question of the most economical combination of
constraints that serves to stabilize a given configuration.
``Economy'' in this case refers to the number of constraints required
to accomplish stabilization.  We are able to establish in the case of
particular interest to us that a nucleosome-like configuration of DNA
is rendered stable by four constraints, and that no fewer constraints
will accomplish this.

The conjecture that underlies the work reported here is that the
portion of DNA that is wrapped around a histone is in a state of
unstable mechanical equilibrium that is rendered mechanically stable
by the constraints associated with the histone spool.  There are
reasons to believe that such a state is desirable.  Imagine a pencil
standing vertically on its point.  While this state will not persist
in the absence of outside influences, it can be sustained without
application of substantial external forces.  In fact, no force at all
is required to guarantee the persistence of this state if the pencil
is exactly vertical.  By the same token, the removal of the constraint
that keeps the vertically balanced pencil from toppling is
accomplished at no energetic cost.  Viewed in this way, the nucleosome
configuration represents a highly efficient stratagem for the local
packing of DNA, in that the histone spool is introduced and removed
with minimal expenditure of resources.

Of course, because DNA is, in its ``naked'' state, highly charged,
electrostatic interactions play an important role in the behavior of
this molecule, both \emph{in vitro} and \emph{in vivo}.  These
interactions are, apparently, key to the condensation of DNA in
prokaryotes \cite{liu}.  Furthermore, electrostatic interactions,
highly screened though they may be, could be important contributors to
the energetics of the DNA-histone interactions.  They are ignored
in this work.  Nevertheless, the elucidation of the stabilization of
configurations within the context of the purely elastic model of DNA
ought to aid in the investigation of similar questions for DNA subject
to the influence of additional interactions and degrees of freedom.

An outline of the paper is as follows.  In Section \ref{sec:hamil} the 
Hamiltonian governing configurations of DNA, modeled as a bent and 
twisted rod, is presented, along with the quadrature solutions to the 
extremum equation for that Hamiltonian.  Section \ref{sec:stab} 
reviews the formulation of linear stability analysis in this case.  A 
key result of the considerations outlined in this section is that any 
non-trivial extremal configuration of bent and twisted DNA will be 
mechanically unstable if the strand is long enough, and if there are 
no constraints on fluctuations.  This means that all extremum 
solutions for an infinite strand of bent and twisted DNA represent 
saddle points of the energy.  Section \ref{sec:constraints} introduces 
the notion of mechanical constraints, particularly those constraints 
associated with the requirement that the straight-line distance 
between two points on the DNA does not change as the segment 
fluctuates.  The way in which this and other constraints are 
mathematically implemented is discussed in this section, and in 
Appendix \ref{app:constraints}.  In section \ref{sec:constraintnumber} 
we address the issue of the number of constraints required to 
stabilize a section of bent and twisted rod against fluctuations about 
an extremum solution.  We find that the minimum number of constraints 
needed to do this is equal to the number of unstable eigenmodes of a 
fluctuation operator introduced in Section \ref{sec:stab}.  Section 
\ref{sec:periodic} contains a discussion of the effects of a periodic 
array of constraints on the stability of an infinitely long section of 
bent and twisted rod.  This discussion provides a lead-in to our 
investigation into the influences required to stabilize the 
nucleosomal configuration of DNA. It is also relevant to the 
stabilizing action of a protein armature on DNA in chromatin.  
Sections \ref{sec:preliminaries} and \ref{sec:stabilization} directly 
address the issue of the stabilization of DNA in the nucleosomal 
configuration.  There are four unstable modes in the case that we 
investigate.  We find that four constraints suffice to counteract 
them.  However, we also find that those constraints must be chosen 
with care.  The effective set reflects the known structure of histone, 
in particular a spiral groove that has been identified.  In our model, 
this groove acts to limit the ability of the DNA wrapped around it to 
slide parallel to the spool's axis.

\section{Hamiltonian and extremum solutions}
\label{sec:hamil}

\begin{figure}
\centerline{\epsfig{file=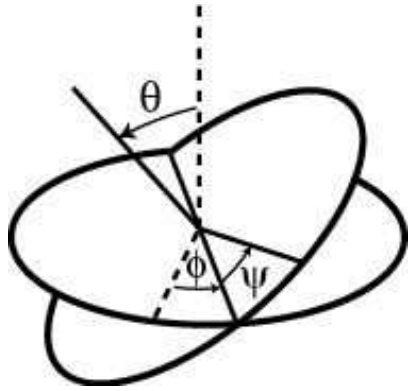,height=2in}}
\caption{The Euler angles, $\theta$, $\phi$ and $\psi$.}
\label{fig:Euler}
\end{figure}

The configuration of the twisted, writhing rod is characterized in
terms of the Euler angles, depicted in Figure \ref{fig:Euler}. We
assume an isotropic rod, characterized by a bending modulus $A$ and a
torsional modulus $C$. The elastic energy of the rod, in terms of the
Euler angles $\theta(s)$, $\phi(s)$ and $\psi(s)$, is given by
\begin{equation}
E_{\rm elastic} = \int \left\{ \frac{A}{2} \left[ \left(\frac{d
\theta(s)}{ds} \right)^{2} + \sin^{2} \theta(s) \left(\frac{d
\phi(s)}{ds}\right)^{2} \right] + \frac{C}{2}\left[ \frac{d
\psi(s)}{ds} + \cos \theta(s) \frac{d \phi(s)}{ds}\right]^{2} \right\}
ds
	\label{energy1}
\end{equation}
Here, $s$ is the arclength along the rod. A way to determine the
equilibrium configurations of the rod is to supplement $E_{\rm
elastic}$ with the term
\begin{equation}
E_{c} = - F \int  \cos \theta(s) ds
	\label{Ec}
\end{equation}
This contribution can either be seen as a Lagrange multiplier that
enforces a given end-to-end distance, or as representing the effect of
tension on the rod.

Finally, in certain cases an additional constraint guarantees
constancy of the end-to-end linking number.  In the case of a rod with
clamped ends, this quantity is given by
\begin{equation}
Lk = -\frac{1}{2 \pi} \int \left(\frac{d \phi(s)}{ds} + \frac{d
\psi(s)}{ds} \right) ds
	\label{link1}
\end{equation}
Note that (\ref{link1}) is an integral over a perfect differential.  
This reflects the topological character of the linking number.  A 
fixed linking number is enforced with the use of a Lagrange 
multiplier.  The quantity to be minimized is, then, the combination
\begin{equation}
E_{\rm elastic} + E_{c} - \lambda Lk
	\label{combo1}
\end{equation}
As the linking number is not a quantity of interest here, we ignore
the final term in (\ref{combo1}).

\begin{figure}
\centerline{\epsfig{file=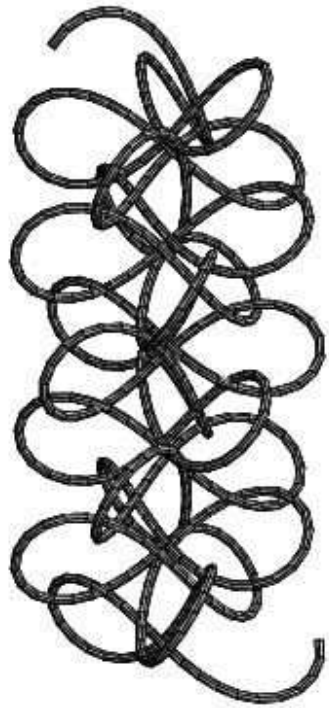,height=4in}}
\caption{The solution to the equation that corresponds to the
nucleosome configuration}
\label{fig:sol}
\end{figure}

The extremum equations have been extensively investigated
\cite{benham}.  They are identical to the equations for the behavior
of the heavy symmetric top.  The connection between top motion and the
deformations of a thin rod was first noted by Kirchhoff, in 1859
\cite{kirc}, and is known as the ``kinetic analogue'' \cite{love}.
Those equations reduce to the following set of three:
\begin{eqnarray}
\frac{d \phi}{ds}&=&\frac{J_{\phi} - J_{\psi} \cos \theta}{A
\sin^{2}\theta} \\
\frac{d \psi}{ds}&=&\frac{J_{\psi}}{C}-\frac{d \phi}{ds} \cos \theta
\\
\sqrt{\frac{A}{2}}\frac{d \theta}{ds}&=&\sqrt{-\frac{\left(J_{\phi}-
J_{\psi} \cos \theta\right)^{2}}{2A \sin^{2} \theta} + E_{0} -F \cos
\theta}
	\label{quad}
\end{eqnarray}
The quantities $J_{\phi}$, $J_{\psi}$ and $E_{0}$ in the above 
equations are integration constants.  The quadrature result for the 
angle $\theta$ is the result of integration of $\delta E_{tot}/\delta 
\theta$.  Defining $u\equiv \cos \theta$, the behavior of solutions 
can be extracted from the equation for $\theta$:
\begin{equation}
ds=\frac{du}{\sqrt{\frac{2\left(1-u^2\right)}{A}
\left(E_0-Fu\right)-
\frac{1}{A^2}\left(J_\phi^2+J_\psi^2-2 J_\phi J_\psi u \right)}} \\
\equiv \frac{du}{\sqrt{\frac{2F}{A}(u-a)(u-b)(u-c)}}
\\
	\label{sol}
\end{equation}
where $c<u<b<a$.  The characteristics of the solutions depend on
quantities $a,b$ and $c$.  The property of solutions have been completely
investigated \cite{fain} for different values of $a,b$ and $c$.  We
focus our attention on those configurations which have the same
configurational form as a segment of DNA in a nucleosome.  For
specific value of $a,b$ and $c$ we have obtained a solution which is
depicted in Figure \ref{fig:sol}.  The reason that we have made the
choice of parameters above has to do with the close visual
relationship between the conformation of the bend and twisted rod as
displayed in Fig.  \ref{fig:sol} and the DNA in a commonly conjectured
form of the 30-nm spiral \cite{chrom}.  We operate here under the
assumption that this configuration is a reasonable representation of
the organization of DNA in this component of chromatin.

\section{Stability determination}
\label{sec:stab}

The stability of an extremum configuration is determined by altering 
the configuration and calculating the change in the quantity that is 
extremized.  In this case, the quantity of interest is the elastic 
energy.  A stable solution is one that minimizes the energy.  If the 
quadratic effect of any small deviation from this solution is to lower 
the energy, then the solution cannot represent stable equilibrium. 
Instead, the configuration is unstable; it is either a 
maximum energy configuration, or a configuration at a saddle point of 
the energy. In the case of the equations for the elastic energy of a 
twisted and bent rod, the second order effect of second order 
fluctuations is obtained by taking second functional derivatives of 
the expression in Eq.  (\ref{energy1}) with respect to the Euler 
angles $\theta$, $\phi$ and $\psi$, and then by setting those Euler 
angles equal to their classical values.  After a bit of reduction, we 
find that the question of the stability of a classical configuration 
can be framed in terms of the spectrum of the following operator 
\cite{closed}
\begin{equation}
-\frac{d^{2}}{ds^{2}} + V(s)
	\label{schrod1}
\end{equation}
where
\begin{equation}
V(s)=-\frac{1}{4} \frac{x(s)}{\left(1-u(s)^{2}\right)^{2}}+
\frac{1}{2}u(s)
	\label{potential1}
\end{equation}
and
\begin{eqnarray}
x(s) &=& \left(2-u(s)\right)
\left(1-u(s)\right)^{2}(a+1)(b+1)(c+1) \nonumber \\ && +
\left(2+u(s)\right)\left(1+u(s)\right)^{2}
(a-1)(1-b)(1-c)
	\label{xdef}
\end{eqnarray}
Here, $u(s)$ is the solution for $\cos \theta(s)$ displayed implicitly
in Eq.  (\ref{sol}).  If all eigenvalues of the operator
(\ref{schrod1}) are positive, then the solution to the classical
equation is stable.  If any is negative, then there are fluctuations
that decrease the energy to a value below its classical value.  Note
that the operator in question resembles the Hamiltonian for a
one-dimensional particle in the potential $V(s)$.  In figure
\ref{fig:poten}, we display this potential for the choice of
parameters $a$, $b$ and $c$ utilized in this investigation.  The
potential, which has been displayed for an extended range of the
arclength, $s$, has the form of the sort of periodic potential
encountered in discussions of electrons in metals and semiconductors.
As in this case, the eigenvalue spectrum consists of bands of
``allowed'' states, separated by ``forbidden regions.''  The question
is whether all or part of any of the bands lie below zero. As it
turns out, this is, indeed, the case.

\begin{figure}
\centerline{\epsfig{file=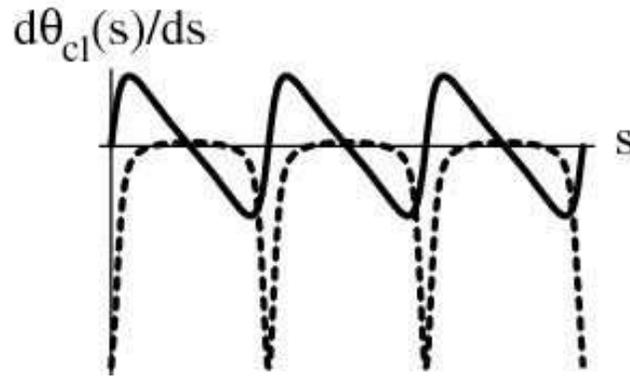,height=2in}}
\caption{The translational mode, $d \theta_{\rm cl}(s)/ds$.  Also
shown, as a dashed curve, is the effective potential, $V(s)$, in the
operator in (\ref{schrod1}).}
\label{fig:translation}
\end{figure}

The reason for this is that one can identify a mode having an
eigenvalue that is \emph{strictly} equal to zero---the
\textbf{translational mode}, equal to the derivative with respect to
$s$ of the classical solution, $\theta_{\rm cl}(s)$.  The existence of
this mode follows from the translational invariance of the extremum
equations, and it is known to play a key role in, for instance, the
question of tunneling between the false and the true vacuum in quantum
field theories \cite{coleman}.  The translational mode in this case is
displayed in Figure \ref{fig:translation}.  Note that this mode is not
spatially uniform, and, in particular, that it possesses nodes.  On
the basis of elementary considerations, one knows that there are, as a
result, solutions to the effective Schr\"{o}dinger's equation
associated with lower---hence negative---eigenvalues.  Figure
\ref{fig:bands} displays the band structure associated with the
potential in Figure \ref{fig:poten}.  In an infinitely long section of
twisted and bent rod that has taken the configuration pictured in
Figure \ref{fig:sol}, there is an infinitely large set of distortions
that will lead to a lowering of the rod's total elastic energy.  In
fact, the reasonable expectation is that these distortions are a route
to interwinding.

\section{The sources and mathematical implementation of
physical constraints}
\label{sec:constraints}

Stabilization of the classical, or extremum, configuration can be
achieved by the introduction of mechanical constraints.  These
constraints are expressed mathematically as the requirement that a 
property of the DNA's configuration does not change under
distortions about the extremum solution.  The physical constraint that
a certain quantity be kept constant translates fairly
straightforwardly into a set of mathematical conditions on the
fluctuation spectrum.  In turn, these mathematical restrictions lead
to a reformulation of the method by which the fluctuation spectrum is
determined.  The reasoning leading from physical constraints to a new
approach to the determination of effective eigenvalues of the linear
fluctuation energy operator is presented in Appendix
\ref{app:constraints}.

Briefly, a constraint on the conformation of a bent and twisted rod
is  expressed mathematically in terms of a condition of the form
\begin{equation}
\int {\cal F} \left( \theta(s), \phi(s), \psi(s) \right) ds = {\rm a \
constant}
	\label{constraint1}
\end{equation}
Under the assumption that Eq.  (\ref{constraint1}) is satisfied for
the extremal configuration, one then expands to the first order in
deviations from the extremal forms of $\theta(s)$, $\phi(s)$ and
$\psi(s)$.  As it turns out, the first order corrections to $\phi(s)$
and $\psi(s)$ are readily expressed in terms of the correction to
$\theta(s)$.  This is because of the simple way in which these two
angles enter into the expression for the total energy in Eq.
(\ref{energy1}).  If we denote by $\gamma(s)$ the displacement of
$\theta(s)$ from its ``classical'' form, the general constraint
equation, (\ref{constraint1}) becomes
\begin{equation}
\int f(\theta_{\rm cl}(s), \phi_{\rm cl}(s), \psi_{\rm cl}(s))
\gamma(s) ds =0
	\label{constraint2}
\end{equation}
where the subscript ``cl'' indicates that the quantity is a solution
to the extremum equations.

For closed configurations, such constraints arise naturally from the
preservation of the topology (i.e.  linking number) of the original
configuration, and from the requirement that the distorted rod
continues to close smoothly on itself.  Here, such considerations do
not necessarily apply, although one might imagine cases in which a
``pinning'' of the ends of a segment forbids any alteration of the
linking number.  Nevertheless, ``boundary conditions'' that result
from physical constraints on the end-points of a given segment do give
rise to mathematical constraints on fluctuations about a given
configuration.  Whether or not boundary conditions will stabilize a
segment of bent and twisted rod against thermally-driven fluctuations
depends on the length of the segment.  If the segment is short enough
compared to the persistence length of the rod, such stabilization is
possible.

Constraints may also be imposed as the result of physical barriers.
For example, imagine that the displacement vector between two points
on the bent and twisted rod is not allowed to vary.  One might imagine
such a constraint being enforced with the use of a stiff, inextensible
``brace'' firmly attached to the rod at the two points in question.
This brace is then immobilized against rotations.  The displacement
vector between the two points is represented as
\begin{equation}
R_{0}=\hat{x} x_{0} + \hat{y}
y_{0} +
\hat{z}z_{0}
	\label{dist1}
\end{equation}
where
    \begin{eqnarray}
x_{0}&=& \int_{s_{1}}^{s_{2}} \sin \theta(s) \cos \phi(s) ds
\\
y_{0}&=& \int_{s_{1}}^{s_{2}} \sin \theta(s) \sin \phi(s) ds
\\
z_{0}&=&\int_{s_{1}}^{s_{2}} \cos \theta(s) ds
	\label{dist2}
\end{eqnarray}
The constancy of each component of this displacement vector is ensured
by a set of three constraints on the deviations of the Euler angle
$\theta(s)$ from its extremum value. If we write
\begin{equation}
\theta(s) = \theta_{cl}(s) = \gamma(s)
	\label{expandtheta}
\end{equation}
then, the following three conditions hold: \begin{enumerate}
\item Constancy of the $x$-component of the displacement vector
\begin{equation}
\delta x=\int_{s_{1}}^{s_{2}}\left\{ u(s) \cos \phi_{cl}(s)
+\sqrt{\frac{F}{2A}}\frac{p_{1}\left(1-u(s)
\right)^{2}+p_{2}\left(1+u(s)
\right)^{2}}{\left(1-u(s)^{2}\right)^{3/2}}{\cal I}_{x}(s) \right\}
\gamma(s) \, ds =0
\label{xconstant}
\end{equation}
where the quantity ${\cal I}_{x}(s)$ is given by
\begin{equation}
{\cal I}_{x}(s) = \int_{s_{1}}^{s}\sqrt{1-u(s')^{2}}\sin \phi_{cl}(s')
\, ds'
	\label{Ixdef}
\end{equation}
\item Constancy of the $y$-component of the displacement vector
\begin{equation}
\delta y=\int_{s_{1}}^{s_{2}}\left\{ u(s) \sin \phi_{cl}(s)
-\sqrt{\frac{F}{2A}}
\frac{p_{1}\left(1-u(s)
\right)^{2}+p_{2}\left(1+u(s)
\right)^{2}}{\left(1-u(s)^{2}\right)^{3/2}}{\cal I}_{y}(s) \right\}
\gamma(s) \, ds =0
	\label{yclosure2}
\end{equation}
where
\begin{equation}
{\cal I}_{y}(s) = \int_{s_{1}}^{s}\sqrt{1-u(s')^{2}}\cos \phi_{cl}(s')
\, ds'
	\label{Iydef}
\end{equation}
\item constancy of the $z$-component of the displacement vector
\begin{equation}
\delta z=\int_{s_{1}}^{s_{2}}\sqrt{1-u(s)^{2}} \gamma(s) \, ds =0
	\label{zclosure2}
\end{equation}
\end{enumerate}
In the above relations
\begin{equation}
p_{1\choose 2}\equiv\left[(c \pm 1)(b \pm 1)(a \pm
1)\right]^{1/2}
	\label{p12}
\end{equation}
As an alternative to the above set of three constraints, one might
imagine that the projection of the displacement vector in a given
direction is held constant, in which case the contstraint is a
linear combination of those constraints:

\begin{equation}
    x_{0}\delta x + y_{0}\delta y + z_{0}\delta z=0
	\label{dist1a}
\end{equation}
As an example of the use of this less restrictive constraint on
fluctuations of the bent and twisted rod, imagine that the brace is
allowed to rotate, but that it remains stiff and inextensible.  Then
the single constraint that holds is that the projection of the
displacement vector along the original direction of the brace is held
fixed.  Thus, physical constraints on the possible contortions of a
strand of DNA translate straightforwardly into mathematical
constraints on the fluctuations of that strand about its ``classical''
configurations.

\section{Number of constraints required to stabilize a configuration}
\label{sec:constraintnumber}

It appears intuitively obvious that the stabilization of a given
configuration, when there are a given number of unstable modes,
requires the imposition of an equal number of constraints.  It is
fairly straightforwardly demonstrated that if there are $n$ unstable
modes, then at least $n$ constraints are required to stabilize them.
To see that this is true, suppose that the operator $L$ has four
negative eigenvalues.  Also, imagine that three constraints have been
imposed, of the form
\begin{equation}
\langle f | \chi_{j} \rangle = \int f(x) \chi_{j}(x) dx =0
	\label{con1}
\end{equation}
for any fluctuation, $f(x)$. Here $1 \le j \le 3$. The four
eigenfunctions having negative eigenvalues will be $\xi_{i}(x)$,
with $1 \le i \le 4$. Let's define
\begin{equation}
g_{ij} \equiv \int \xi_{i}(x) \chi_{j}(x) dx
	\label{gdef}
\end{equation}
Now, take a fluctuation that is of the form
\begin{equation}
f(x) = \sum_{i=1}^{4} a_{i} \xi_{i}(x)
	\label{ftrial}
\end{equation}
The three constraint equations are of the form
\begin{equation}
\sum_{i=1}^{4} a_{i}g_{ij}=0
	\label{con2}
\end{equation}
These are three equations in the four unknowns $a_{i}$.  We'll assume
that not all $g_{ij}$'s are equal to zero for any $i$.  Then, it is
possible to set one of the $a_{i}$'s equal to one.  The equations
reduce to three linear, inhomogeneous, equations in three unknowns.
Unless there is some degeneracy, it will be possible to find a
solution to those equations.  This means that a function of the form
(\ref{ftrial}) will obey the constraints.  Furthermore the expectation
value
\begin{equation}
\langle f|L|f \rangle = \int f(x) L(x,x^{\prime}) f(x^{\prime}) dx \
dx^{\prime}
	\label{expect1}
\end{equation}
will be given by
\begin{equation}
\langle f|L|f \rangle  = \sum_{i=1}^{4} a_{i}^{2} \lambda_{i}
	\label{expect3}
\end{equation}
Given that the four $\lambda_{i}$'s in the sum are all negative, we
have a fluctuation for which the expectation value of the linear
operator $L$ is negative.

It is, thus, clear that three constraints do not suffice to stabilize
a classical configuration against fluctuation when there are four
unstable modes.  This conclusion generalizes straightforwardly to the
case of $n$ unstable modes and $m<n$ constraints.  On the other hand,
$n$ constraints may \underline{or may not} prove sufficient to guarantee
stability.  Consider, for example the case of a single instability.
Let the unstable mode be $\xi_{0}(x)$. If the single constraint
requires that all fluctuations be orthogonal to $\chi(x)$, then the
equation satisfied by the eigenvalues, $\lambda$ of the constrained
fluctuation operator is
\begin{equation}
\sum_{i}\frac{\left(\int \xi_{i}(x) \chi(x)
dx\right)^{2}}{\lambda_{i} - \lambda} =0
	\label{fluceq}
\end{equation}
It is straightforwardly verified that solutions of this equation lie
between consecutive eigenvalues, $\lambda_{i}$ of the unconstrained
fluctuation operator.  Thus, the lowest allowed eigenvalue of the
constrained operator lies above the lowest eigenalue in the
unconstrained system. However, it also lies below the next-lowest
unconstrained-system eigenvalue.

To see that stabilization may or may not occur in this case, we
consider two particular subsets of the many possible alternatives for
the function $\chi(x)$.  First, imagine that $\chi(x) \propto
\xi_{0}(x)$.  Then the constraint entirely eliminates the unstable
mode and stability is guaranteed.  On the other hand, suppose that
$\chi(x) \propto \xi_{i}(x)$ with $i \neq 0$.  Then, a stable mode is
eliminated, and the unrestrained unstable mode contributes to the
fluctuation spectrum.  The instability is entirely unaffected.

\section{The effects of a periodic array of constraints}
\label{sec:periodic}

In the case of DNA confined to the nucleus of a cell, it is widely
conjectured that the packing of DNA is accomplished with the use of a
hierarchical organization of the long strands that constitutes the
genome.  Given that this organization will not be a mechanically
stable structure, at least within the bent-and-twisted rod model of
DNA, some constraining mechanism is required.  The histone spools
provide one set of constraints, but these operate on the lowest level
of organization.  It is possible that a protein ``armature''
provides the necessary stabilization at higher levels.  Here, we
discuss the implications of the kinds of mathematical constraints on
fluctuations about unstable mechanical equilibrium that one can
reasonably associate with the mechanical influence of this mechanism
for stabilization.

In this context, we focus on the case of a long strand of distorted
DNA, or, equivalently, a long section of bent and twisted rod.  Here,
the set of fluctuations that lowers the energy of the unstabilized
configuration is quite large.  This implies the need for a large
number of constraints.  When the strand is infinitely long, and the
number of unstable mechanical modes is infinite, then an infinite
number of constraints is required. We will look here at the
stabilizing effect of a periodic array of constraints.

The operator $L$ controlling the stability of the equilibrium
configuration of a long segment of the bent and twisted rod has the
form
\begin{equation}
-\frac{d^{2}}{dx^{2}} +V(x)
	\label{periodic1}
\end{equation}
Where the potential term, $V(x)$, is periodic, in that
\begin{equation}
V(x+a)=V(x)
	\label{periodic2}
\end{equation}
According to Floquet's theorem, the eigenfunctions of the above
operator are of the form
\begin{equation}
\phi_{n,k}(x) = e^{ikx}C_{n,k}(x)
	\label{Floquet1}
\end{equation}
where $k$, called the crystal momentum in solid state physics, is
confined to a Brillouin zone.  The most convenient Brillouin zone for
our purposes is $-\pi/a \le k < \pi/a$.  The function $C_{n}(x)$ is
periodic in $x$, in that
\begin{equation}
C_{n,k}(x+a) = C_{n,k}(x)
	\label{Floquet2}
\end{equation}
The integer $n$ is called the band index. The eigenvalue of this
eigenfunction is also indexed by the crystal momentum and the band
index, i.e. $E_{n,k}$.

If a periodic array of constraints is imposed, in that we require all
fluctuations to be equal to zero at $x=b+ma$, with $m$ an integer
and $-\infty < m < \infty$, then the requirement that the determinant
is equal to zero translates into the requirement that the following
product is equal to zero:
\begin{equation}
\prod_{k=-\pi/a}^{\pi/a}{\cal F}(k)
	\label{product1}
\end{equation}
where
\begin{equation}
{\cal F}(k) =
\sum_{n}\frac{\left|C_{n,k}(b)\right|^{2}}{E_{n,k}-\lambda}
	\label{product2}
\end{equation}
This tells us that for every value of $k$, the lowest value of
$\lambda$ corresponding to a fluctuation lies between the lowest value
of $E_{n,k}$, as a function of the band index $n$ and the next lowest
value of that eigenvalue. Suppose we impose two constraints, by
requiring that the fluctuations are zero at $x=b_{1}+ma$ and $x=b_{2}+ma$.
Then, the determinant will consist of a product of terms of the form
\begin{equation}
\sum_{n_{a}>n_{b}}\frac{\left(C_{n_{a},k}(b_{1})C_{n_{b},k}( b_{2}) -
C_{n_{a},k}(b_{2})C_{n_{b},k}(b_{1}) \right)^{2}}{\left(E_{n_{a},k}-
\lambda \right) \left(E_{n_{b},k} - \lambda \right)}
	\label{product3}
\end{equation}
In this case, it is possible that the lowest solution of the
characteristic equation will lie even higher than when there is only
one constraint per period.

\begin{figure}
\centerline{\epsfig{file=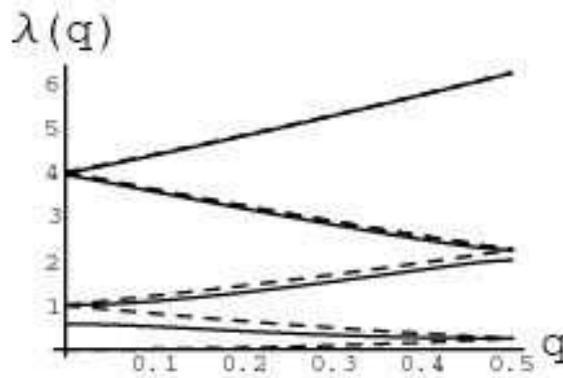,height=2in}}
\caption{Alteration of the band structure as the result of a periodic
array of constraints.  The dotted curves are the bands in the absence
of constraints.  The solid curves represent the influence of
constraints.}
\label{fig:bands1}
\end{figure}

As an indication of the effect of an array of constraints, we
consider the case of the eigenstates of the ``periodic'' potential
that is equal to zero everywhere. As is well-known, one can imagine a
one-dimensional Brillouin zone of fixed width. The dispersion
relation can then be expressed in terms of a series of curves in which
the ``crystal momentum'' is restricted to this zone. There are no band
gaps, but otherwise the bands are well-behaved. Now, one is interested
in the expectation values of the operator
\begin{equation}
L_{0} = - d^{2}/dx^{2}
	\label{L0def}
\end{equation}
Imagine that the lattice spacing is one, and take for the function
to which fluctuations are orthogonal a gaussian of the form
\begin{equation}
\chi(x) = e^{-x^{2}}
	\label{phifunc}
\end{equation}
The array of functions are copies of (\ref{phifunc}) centered about
the points $x=\pm 1, \pm 2 , \pm 3\ldots$ The solution to the
equation setting (\ref{product1}) equal to zero is graphed in Figure
\ref{fig:bands1}. Also shown in that figure are the ``bands'' of the
unconstrained operator. Note that the allowed values of $\lambda$ for
fixed crystal momentum, $q$, lie between successive bands. This is a
general feature of array of constraints.

\begin{figure}
\centerline{\epsfig{file=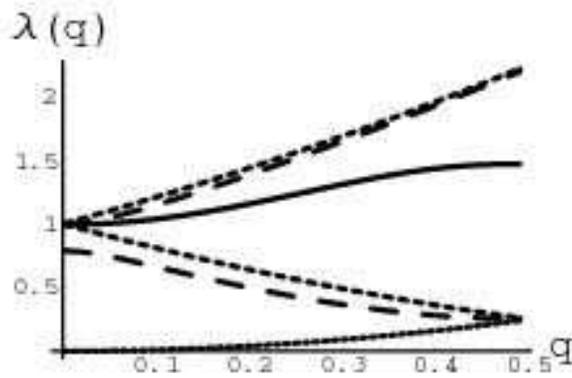,height=2in}}
\caption{Alteration of the band structure as the result of a periodic
array of constraints, when there are two constraints per period.  Here
the solid curve is the result for $\lambda(q)$ when there are two
constraints per period.  The dashed curve represents the influence of
one constraint, and the dotted curves are the bands in the absence of
constraints.  The solid curves represent the influence of
constraints.}
\label{fig:bands2}
\end{figure}

The effects of a pair of constraints in every period is illustrated in
Figure \ref{fig:bands2}, where the values of the $\lambda$ for both
one and two constraints per period are compared with the
$\lambda$-versus-$q$ relationship for the unconstrained operator.

\begin{figure}
\centerline{\epsfig{file=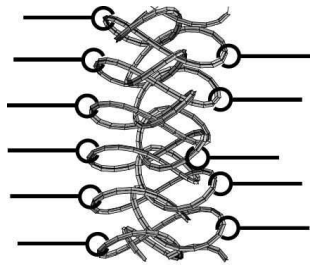,height=2in}}
\caption{A schematic representation of a non-trivial supercoiled
configuration of DNA stabilized by a protein armature.}
\label{fig:armature}
\end{figure}

This brings us to the way in which a physical armature, in the form of
a protein scaffolding, can act to stabilize a nontrivially
supercoiled DNA configuration. We imagine a configuration as depicted
in Figure \ref{fig:armature}. The contacts between the DNA and the
armature will stabilize the DNA against fluctuations.

\section{The case of a nucleosome: preliminaries}
\label{sec:preliminaries}

\begin{figure}
\centerline{\epsfig{file=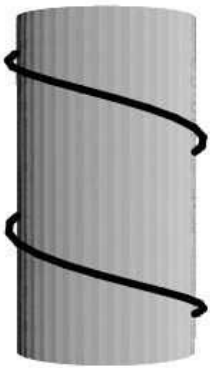,height=2in}}
\caption{Schematic of the nucleosome.}
\label{fig:nuc}
\end{figure}

In the nucleosome configuration a segment of DNA wraps around a
collection of proteins known as a histone \cite{thomas}.  In the
schematic depiction of the nucleosome, the DNA segment is represented
as a spiral surrounding a cylinder.  See Figure \ref{fig:nuc}.  As a
first step in our investigation of the stability of the spatial
configuration of the segment of DNA that participates in the
nucleosome we will look at the stability of the spiral solution to the
energy extremum equations for a bent and twisted rod.

Now, the spiral is a special, limiting case of the solutions to the
classical equation for $u(s) = \cos \theta(s)$.  In this solution, the
equality $b=c$ holds, and $u(s)$, which lies between those two
parameters in the classical solution is, thus, fixed at their common
value, which we henceforth will call $b$.  The complete determination
of the solution requires that we set the parameter $a$ and choose
signs in (\ref{p12}).  We find that there are four possibilities for
the solution, corresponding to the four choices of the two signs.  Two
of the solutions are for a left-handed spiral, and in the other two
the spiral is right-handed.  In the case of the nucleosome, DNA is
wrapped around the histone spool in a left-handed spiral.  Given the
sense of the helical solution, there the two alternative solutions are
spirals that the arclength of a single turn of which is either greater
or less than $2 \pi \sqrt{F/A}$.  The quantity $\sqrt{F/A}$ is the
persistence length of the rod, and the only intrinsic length scale in
this system.  If we rescale arclengths so that they are expressed in
units of $\sqrt{F/A}$, then the rate of change of the Euler angle
$\phi(s)$ in this classical solution is given by
\begin{equation}
\frac{d \phi(s)}{ds} = \frac{\sqrt{a+1} \pm \sqrt{a-1}}{\sqrt{2}}
	\label{pitch1}
\end{equation}
The diameter of the cylindrical region encircled by the helical
solution is given by
\begin{equation}
\frac{2 \sqrt{2}}{\sqrt{a+1} \pm \sqrt{a-1}} \sin \theta =
\sqrt{1-b^{2}} \sqrt{2} \left(\sqrt{a+1} \mp \sqrt{a-1} \right)
	\label{diam}
\end{equation}
while the distance between successive turns of the helix, measured
along the direction parallel to the cylinder's axis, is given by
\begin{equation}
\frac{2 \pi \sqrt{2}}{\sqrt{a+1} \pm \sqrt{a-1}} \cos \theta =
\sqrt{2}\pi \left(\sqrt{a+1} \mp \sqrt{a-1} \right) b
	\label{dist}
\end{equation}

The linear equation, the eigenvalues of which yield the energies of
fluctuations about the classical solution, is
\begin{equation}
-\frac{d^{2} \Phi(s)}{ds^{2}} + (b-a) \Phi(s) = \lambda \Phi(s)
\label{newfluc}
\end{equation}
The parameter $a$ must be greater than one to ensure real solutions to
the classical equations, while $b = \cos \theta$ lies between 1 and
-1.  If the length of the spiral is allowed to become infinite, then,
in the absence of constraints, there are an infinite number of
solutions to Eq.  (\ref{newfluc}) with negative values of $\lambda$.

\begin{figure}
\centerline{\epsfig{file=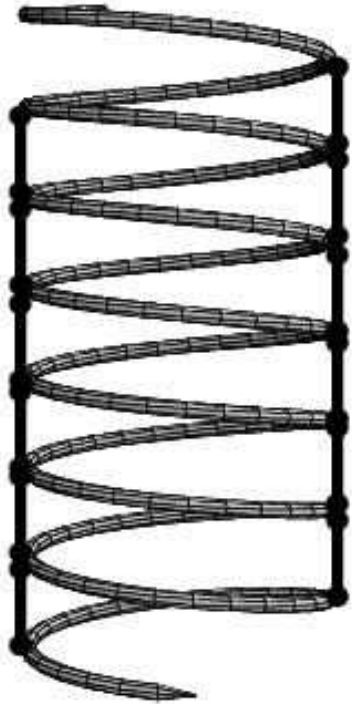,height=4in}}
\caption{The constrained spiral}
\label{fig:conspi}
\end{figure}

Eq.  (\ref{newfluc}) is just the kind of equation for which one might
envision stabilization as the result of the imposition of a regular
array of constraints.  Here, we ask what physical constraints will
have the effect of stabilizing the extended helix against
fluctuations.  One possibility is depicted in Figure \ref{fig:conspi}.
The vertical dark lines in the figure represent rods that enforce a
fixed spacing between a point on the spiral and the point immediately
above or below it.  As shown in the figure, there are two ``lines'' of
these rods, on opposite sides of the spiral.  As we will see, this
arrangement proves sufficient to stabilize a family of spirals against
fluctuations.

\begin{figure}
\centerline{\epsfig{file=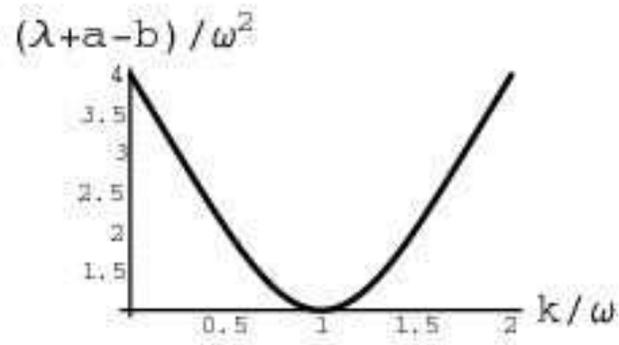,height=2in}}
\caption{The eigenvalue of the constrained fluctuation operator in the
case of interest here. }
\label{fig:conspigraph}
\end{figure}

The requirement that the distance between a point on the spiral and a
point separated from it by a single turn of the spiral translates
into the following requirement on a fluctuation, $\Phi(s)$
\begin{equation}
\int_{s_{0}}^{s_{0} + \tau} \phi(s) ds =0
	\label{consp1}
\end{equation}
Here, $s_{0}$ is the location of the first point along the spiral,
while $\tau$ is the ``period'' of the spiral, the backbone distance
from a point on it to the point one turn of the spiral subsequent. In
this case
\begin{equation}
\tau = \frac{\sqrt{2} }{\sqrt{a+1} \pm \sqrt{a-1}}2 \pi
	\label{taudef}
\end{equation}
We will henceforth take the sign in (\ref{taudef}) to be the upper
one, corresponding to the more tightly wound of the two branches.
Then, the periodic set of constraints indicated in Figure \ref{fig:conspi}
leads to the following equation for the eigenvalues of the constrained
spiral
\begin{equation}
\sum_{n=-\infty}^{\infty} \frac{1}{\left( k+2 n
\omega\right)^{2}\left[\left( k+2 n
\omega\right)^{2}- \left( \lambda +a-b \right) \right]} =0
	\label{conspi2}
\end{equation}
where
\begin{equation}
\omega = \frac{2 \pi}{\tau}
	\label{omegadef}
\end{equation}
This equation  is a specific realization of (\ref{product2}), in which
factors that are independent of the summation variable $n$ have been
ommitted. The sum in (\ref{conspi2}) can be performed with the use of
contour integration. The equation that results is
\begin{equation}
\frac{\pi \,\omega \,\cot (\frac{k\,\pi }{2\,\omega } -
          \frac{\pi \,{\sqrt{a - b + \lambda }}}{2\,\omega }) -
       \pi \,\omega \,\cot (\frac{k\,\pi }{2\,\omega } +
          \frac{\pi \,{\sqrt{a - b + \lambda }}}{2\,\omega }) -
       {\pi }^2\,{\sqrt{a - b + \lambda }}\,
        {\csc (\frac{k\,\pi }{2\,\omega })}^2}{4\,
       {\left( a - b + \lambda  \right) }^{\frac{3}{2}}\,
       {\omega }^2} =0
	\label{conspi3}
\end{equation}
The minimum value of the $\lambda$ that solves this equation
corresponds to $k=\pm \omega$, at which point
\begin{eqnarray}
\lambda &=& -a+b+\omega^{2} \nonumber \\
&=& -a+b + \left(\frac{\sqrt{a+1} + \sqrt{a-1}}{\sqrt{2}}
\right)^{2} \nonumber \\
&=& b+\sqrt{a^{2}-1}
\label{conspi4}
\end{eqnarray}
A graph of $(\lambda +a-b)/\omega^{2}$ as a function of $k/\omega$ is
shown in Figure \ref{fig:conspigraph}. The constraints depicted in
Figure \ref{fig:conspi} will keep the spiral in place against thermal
fluctuations.

\section{Stabilization of a single nucleosome in a 30 nm-spiral-like
array}
\label{sec:stabilization}

\begin{figure}
\centerline{\epsfig{file=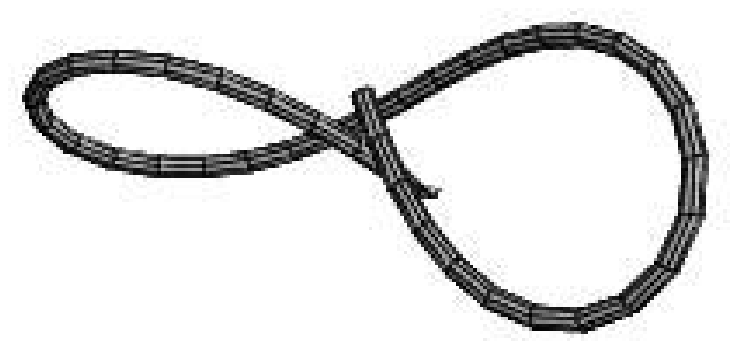,height=2in}}
\caption{The portion of the compound spiral in Fig. \ref{fig:sol} that
corresponds to a nucleosome in the 30 nm spiral.}
\label{fig:portion}
\end{figure}

Here, we take the point of view that there is merit to the notion of
an organized and orderly array of nucleosomes in the 30 nanometer
spiral, and we search for this order in the solution to the energy
extremum equations for a bent and twisted rod. Interestingly,
solutions that mimic a conjectured form of this higher order
structure can be found. One such solution is depicted in Figure
\ref{fig:sol}.  As previously noted, this solution bears a visual
relationship to the coiling of DNA in a conjectured form of the
higher order structure known as the 30 nm spiral. The specific values
of the parameters $a$, $b$ and $c$ that generate this configuration
are
\begin{eqnarray}
a & = & 1
	\label{aval}  \\
b & = & 0.656009130822
	\label{bval}  \\
c & = & -0.85
	\label{cval}
\end{eqnarray}
In this study, we focus on a particular portion of this structure,
corresponding to two loops in the distorted DNA strand. This portion
is illustrated in Figure \ref{fig:portion}. Note that the loops are
not compact as in the standard picture of a nucleosome. The case here
is a bit figurative, as we are interested in the notion of
organization on a larger scale as envisioned in some versions of the
30 nm spiral.

The ``bare'' stability of the two-loop portion of DNA was calculated
by assuming that fluctuations were consistent with free boundary
conditions, in which the slope of the fluctuations in the angle,
$\theta(s)$ is set equal to zero at the two ends of the DNA segment.
With these boundary conditions, we find that there are four unstable
modes of the fluctuation operator (\ref{schrod1}), with potential
$V(s)$ as given by (\ref{potential1}).  The eigenfunctions associated
with those fluctuations are shown in Figure \ref{fig:unstable}.  In
line with the discussion above, this implies the need for at least
four constraints on fluctuations of the segment of DNA that is wrapped
about the histone in this configuration.  The construction of these
eigenfunctions required an elaboration of the integration method that
we generally utilized to find the solution of the linear second order
equation that governs fluctuations about extremal solutions.  This
elaboration is discussed in Appendix \ref{app:transfer}.

We choose to assume that the histone provides constraints in the most
``efficient'' manner, that is, that number of constraints that follow
from the presence of the histone does not exceed the minimum number
required to guarantee stability of the nucleosome configuration.
Histones keep the two loops close to each other and limit the
arbitrary fluctuations of two loops with respect to each other.  With
this in mind, we started by fixing the distance between two different
points on the segment of DNA which wraps around the histone octomer.
As shown in the previous section, at least four constraint functions
are required to stabilize the nucleosome structure.

Our strategy is to construct four constraint functions, each
associated with fixing a different distance on a segment of DNA in
Nucleosomes.  We are then faced with the problem of solving for zeros
of the determinant of the matrix $G_{kl}$.  This matrix is defined in
Appendix \ref{app:constraints}, in Eq.  (\ref{extrem3}). The
operator ${\cal L}$, is given here by
\begin{equation}
{\cal L} = \frac{\Phi(s) \Phi(L-s)}{\Phi^{\prime}(L)}
	\label{prop}
\end{equation}
Here, $\Phi(s)$ is an eigenvalue of the operator (\ref{schrod1}) that
has the property $\Phi^{\prime}(0)=0$.  The quantity $L$ is the total
arclength of the nucleosomal segment.

As an initial attempt, we fixed four ``diagonal'' distances between
two loops.  We fixed these distances only in $x$-$y$ plane.  We assume
that histone has a distorted cylindrical shape and this way we fixed
the radius of cylinder in four different places.  With this set of
constraints, the DNA segment has some freedom to move vertically as
long as it is wrapped around the histone spool.  In this case, we
found that constraints only removed two negative eigenvalues and the
system remains mechanically unstable.  We then tried quite a few set
of constraints related to keeping the segment of DNA loosely on the
histone octomer.  A few of these sets of constraints can be seen in
the Figure \ref{fig:unstable}.  None of these sets of constraints was
able to eliminate all negative eigenvalues.  An example of the
determinant $|G_{kl}|$, defined in Appendix \ref{app:constraints},
associated with one of the sets of inadequate constraints is shown in
Figure \ref{fig:baddet}.

As indicated by the brief account above, the task of constructing such
constraints is by no means trivial.  Four constraints chosen at
random, will not, in our experience prove adequate to the task of
stabilizing the DNA segment against fluctuations.  To understand the
mechanism of removing of a negative eigenvalue better, we constructed
a five by five matrix with the same eigenvalues as the five lowest
eigenvalues of our problem.  We let the computer choose four
constraints randomly and ran the program many times.  We were not able
to see even one case in which the constraints remove the four negative
eigenvalues.  The distance between the third eigenvalue and fourth (as
shown in the picture) is very large compared to distance between other
eigenvalues.  As a result it is not at all easy to find a set of
constraints that eliminates all negative eigenvalues.

\begin{figure}
\centerline{\epsfig{file=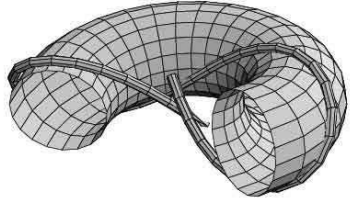,height=2in}}
\caption{The nuclesome configuration in this case. Note the the
``histone'' is in the form of a curved cylinder. The curvature is so
as to fit into the two-loop structure that we investigate here. }
\label{fig:newhistone}
\end{figure}

\begin{figure}
\centerline{\epsfig{file=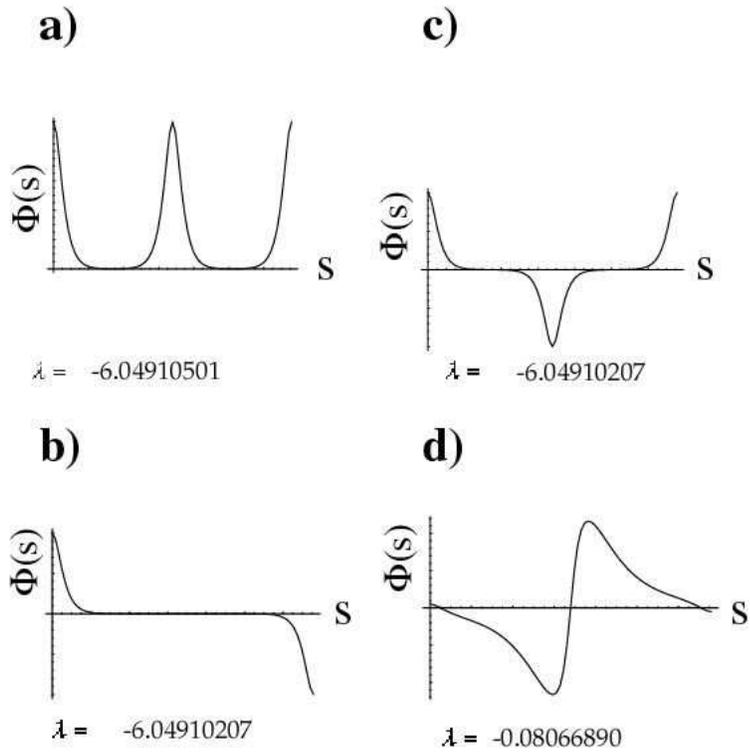,height=4in}}
\caption{The four unstable eigenvalues of the fluctuation operator
for the ``nucleosome'' configuration illustrated in Figure
\ref{fig:newhistone}. The associated eigenvalues are shown immediately
below the graphs.}
\label{fig:unstable}
\end{figure}

\begin{figure}
\centerline{\epsfig{file=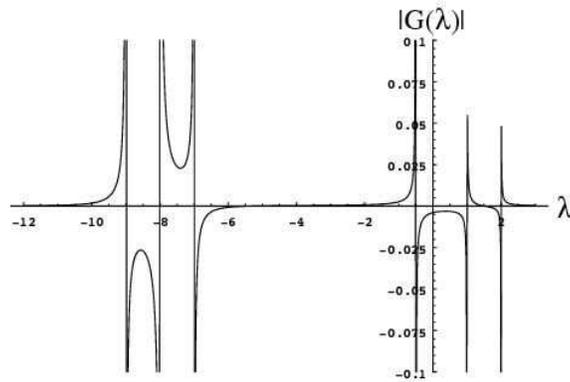,height=2in}}
\caption{A determinant associated with one of the constraints that
does not stabilize the nuclesome configuration illustrated in Figure
\ref{fig:newhistone}. Note that this determinant passes through zero
as a function of $\lambda$ for $\lambda <0$. Recall that zeros of the
determinant are proportional to energy eigenvalues of the fluctuation
operator.}
\label{fig:baddet}
\end{figure}

\begin{figure}
\centerline{\epsfig{file=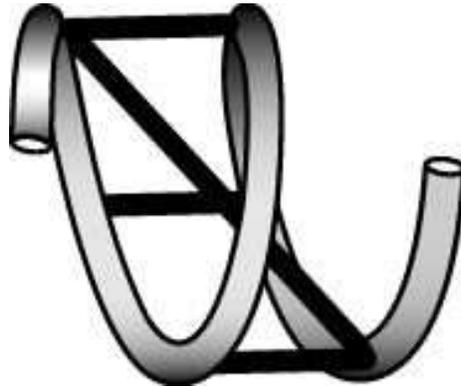,height=2in}}
\caption{A figurative version of the actual set of constraints that
were utilized in this set of calculations. In this picture, the
``histone'' has been straightened out to resemble a the cylindrical
shape that it actually takes. }
\label{fig:newhist}
\end{figure}

In the end, consideration of the detailed structure of the nucleosome,
and a knowledge of the nature of the periodic constraints that
stabilize a long spiral of DNA led to four constraints that give rise
to mechanical stability \cite{note1}.  Three of the four constraints
corresponded to rods that stabilize the segment against motion
parallel to the (curved) axis of the histone, and the fourth is in the
form of a ``diagonal'' strut, reaching nearly across the double-looped
segment.  In more detail, the constraints correspond to rigid, but
hinged, rods that join points in the nucleosome-like segment as
follows: \begin{enumerate} \item The diagonal strut reaches from a
quarter of the way in the first loop, to a quarter of the way from the
end of the second loop.  This is the long, diagonal support
illustrated in Figure \ref{fig:newhist}.  Note that the picture of the
nucleosome here is figurative, in that the ``real'' histone, as shown
in Figure \ref{fig:newhistone}, has a curved axis, so as to fit into
the loops of the nucleosomal DNA.

\item The second support runs parallel to the axis of the histone,
from a quarter in the first loop to a quarter in the second. This is
the topmost horizontal support in Figure \ref{fig:newhist}.

\item The third support, also parallel to the axis of the histone
extends from the halfway point of the first loop to the halfway point
of the second loop.

\item Finally, the fourth support, which, like the second and third
ones, runs parallel to the histone's axis, joins the point three
quarters of the way into the first loop to the point a quarter of the
way into the second loop from the opposite end. This is the bottom
horizontal support in Figure \ref{fig:newhist}.

\end{enumerate}

\begin{figure}
\centerline{\epsfig{file=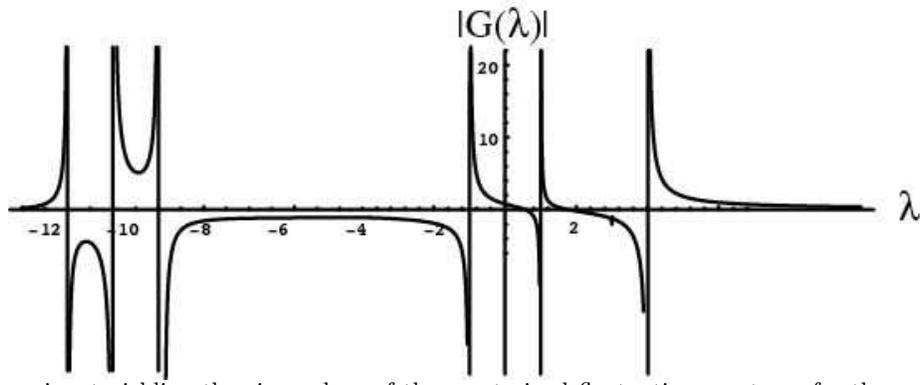,height=2in}}
\caption{The determinant yielding the eigenvalues of the constrained
fluctuation spectrum for the case of the constraints on the two loops
in Figure \ref{fig:portion} that are described in Section
\ref{sec:stabilization}.  Note that this determinant as function of
$\lambda$ does not pass through zero for any $\lambda <0$.  Given that
zeroes of the determinant are proportional to energy eigenvalues of
the constrained fluctuation operator, we are assured that the
nuclesome configuration is stabilized against mechanical fluctuations.}
\label{fig:detgraph}
\end{figure}

The equation for eigenvalues of the fluctuation spectrum, now has the
form of the characteristic equation of the appropriate version of the
matrix $G_{kl}$ displayed in Eq. (\ref{extrem3}). The determinant of
this matrix, as a function of the eigenvalue parameter, $\lambda$, is
shown in Figure \ref{fig:detgraph}. The zeros of the determinant occur
at the eigenvalues of the constrained fluctuation spectrum. We note
that there are no negative roots for negative values of $\lambda$.
The poles  that appear in the plot lie at the locations of the
eigenvalues of the unconstrained spectrum. The four negative energies
are readily identified in the figure. It is worth noting that the
stabilization leaves the segment with a positive eigenvalue that lies
close to zero. In other words the four constraints that were utilized
were adequate to achieve mechanical stability, but only barely so.

As noted above, the choice of the four constraints that led to
stability of the nucleosomal configuration was guided by known
properties of the histone octamer.  A variety of investigations has
revealed the existence of a spiral ``trough'' in the surface of the
histone \cite{luger,arents1,arents2}.  Such a trough will act to
constrain wrapped DNA against movement along the surface of the
histone spool that is parallel to that spool's axis.  In particular,
the section of DNA that is wrapped about the histone spool will not be
allowed to move in such a way as to alter the distance between
adjacent coils, when that distance is measured along a direction
parallel to the spool axis.  In addition, we were guided by the
results reported in Section \ref{sec:preliminaries}, in which it was
demonstrated that an extended spiral is stabilized by a periodic array
of constraints equivalent to a set of rigid, but hinged, rods running
separating consecutive turns of the spiral as indicated in Figure
\ref{fig:conspi}.

 From investigations of the chemical electrostatic and conformational
structure of the histone octamer, it is clear that the points of
contact between the histone and the DNA wrapped around it exceed the
minimal number that, according to our results, stabilize the DNA
segment against mechanical instabilities.  However, it is satisfying
that a ``minimal'' set of constraints will also do the job.  The
significance of this result for the mechanics and biology of the
nuclesome configuration remains to be worked out.  Nevertheless, it
has long been known that a few points of contact between DNA and the
histone spool suffice to stabilize the nucleosome \cite{klug}.  We
believe that issues of optimal efficiency will prove relevant in
discussions of the nuclesome in eukaryotic chromatin.

\section*{Acknowledgements}

The authors acknowledge helpful discussions with W. M. Gelbart, K.-K. Loh,
and V. Oganesyan.

\begin{appendix}

\section{The influence of constraints on the fluctuation spectrum:
general results}
\label{app:constraints}

The mathematical effects of constraints on the fluctuation spectrum of
the operator (\ref{schrod1}) are readily expressed in terms of the
roots of a determinant. Here, we outline the way in which this
formulation of the stability investigation is arrived at. The
discussion in this section has appeared before \cite{closed}. It is
repeated here for the convenience of the reader.

The investigation of the stability of a solution to an Euler-Lagrange
equation, such as the one relevant to the configurations of interest
to us here can be framed in terms of the eigenvalue spectrum of a linear
operator. This, in turn, can be recast in terms of the problem of
finding extremal values for the expectation value
\begin{equation}
\langle \xi |{\cal L}| \xi \rangle
    	\label{exp1}
\end{equation}
where ${\cal L}$ is the linear operator.  In the case at hand, $L$ is
the operator in (\ref{schrod1}).  The constraints are equivalent to
requiring that the $\xi$ between which the operator is sandwiched is
orthogonal to a set of $m$ $\chi$'s.  There is also the constraint on
the absolute magnitude of $\xi$.  The constraints are, then of the
form
\begin{eqnarray}
\langle \xi|\xi \rangle & = & 1
	\label{normalization1}  \\
\langle \xi | \chi_{l}\rangle & = & 0
	\label{orthogonal1}
\end{eqnarray}
In Eq. (\ref{orthogonal1}), the index $l$ runs from 1 to $m$. The
equation for the extremum of the quadratic form (\ref{exp1}), subject
to the constraints (\ref{normalization1}) and (\ref{orthogonal1}),
takes the form
\begin{equation}
{\cal L} |\xi\rangle = \lambda| \xi \rangle +
\sum_{l=1}^{m}\Lambda_{l}| \chi_{l} \rangle
	\label{extrem1}
\end{equation}
The coefficients $\lambda$ and $\Lambda_{l}$ are Lagrange multipliers,
which enforce the constraints to which the system is subject. The
solution to the above equation is
\begin{equation}
|\xi\rangle = \sum_{l=1}^{m}\frac{\Lambda_{l}}{{\cal L}-
\lambda}|\chi_{l}\rangle
	\label{extrem2}
\end{equation}
The Lagrange multipliers $\Lambda_{l}$ must now be adjusted to ensure
the orthogonality requirements. These requirements are of the form
\begin{eqnarray}
0&=&\sum_{l=1}^{m}\Lambda_{l}\langle
\chi_{k}|\frac{1}{{\cal L}-\lambda}|\chi_{l}\rangle \nonumber \\
&\equiv& G_{kl}\Lambda_{l}
	\label{extrem3}
\end{eqnarray}
This set of $m$ equations for the Lagrange multipliers $\Lambda_{l}$
has non-trivial solutions only if the determinant of the $m\times m$
matrix $G$ is zero. The equation $\left|G_{jk}\right|=0$ represents a
condition on the parameter $\lambda$.

Now, given a solution to Eq. (\ref{extrem3}), we take the expectation
value $\langle \xi {\cal L} \xi \rangle$. Substituting from the right hand
side of Eq. (\ref{extrem3}), we find for this expectation value
\begin{eqnarray}
\sum_{l=1}^{m}\langle \xi |{\cal L} \frac{\Lambda_{l}}{{\cal L}- \lambda}
|\chi_{l}\rangle &=& \sum_{l=1}^{m}\langle \xi |\left( {\cal L}- \lambda
\right) \frac{\Lambda_{l}}{{\cal L}-\lambda}| \chi_{l}\rangle +
\lambda\sum_{l=1}^{m} \langle \xi |\frac{\Lambda_{l}}{{\cal L}-\lambda}
\chi_{l}|\rangle \nonumber \\
&=& \sum_{l=1}^{m}\Lambda_{l}\langle \xi |\chi_{l} \rangle + \lambda
\langle \xi | \xi \rangle \nonumber \\
&=& \lambda
	\label{expect2}
\end{eqnarray}
In Eq. (\ref{expect2}) we have made use of the orthogonality of $\xi$
to the $\chi_{l}$'s. We are also assuming that the function $\xi$ is
normalized.  Thus, in  solving for the value of $\lambda$ that satisfies
Eq. (\ref{extrem3}) we are also determining the effective values of the
eigenvalues of the constrained problem.

\section{Transfer Matrix}
\label{app:transfer}

\begin{figure}
\centerline{\epsfig{file=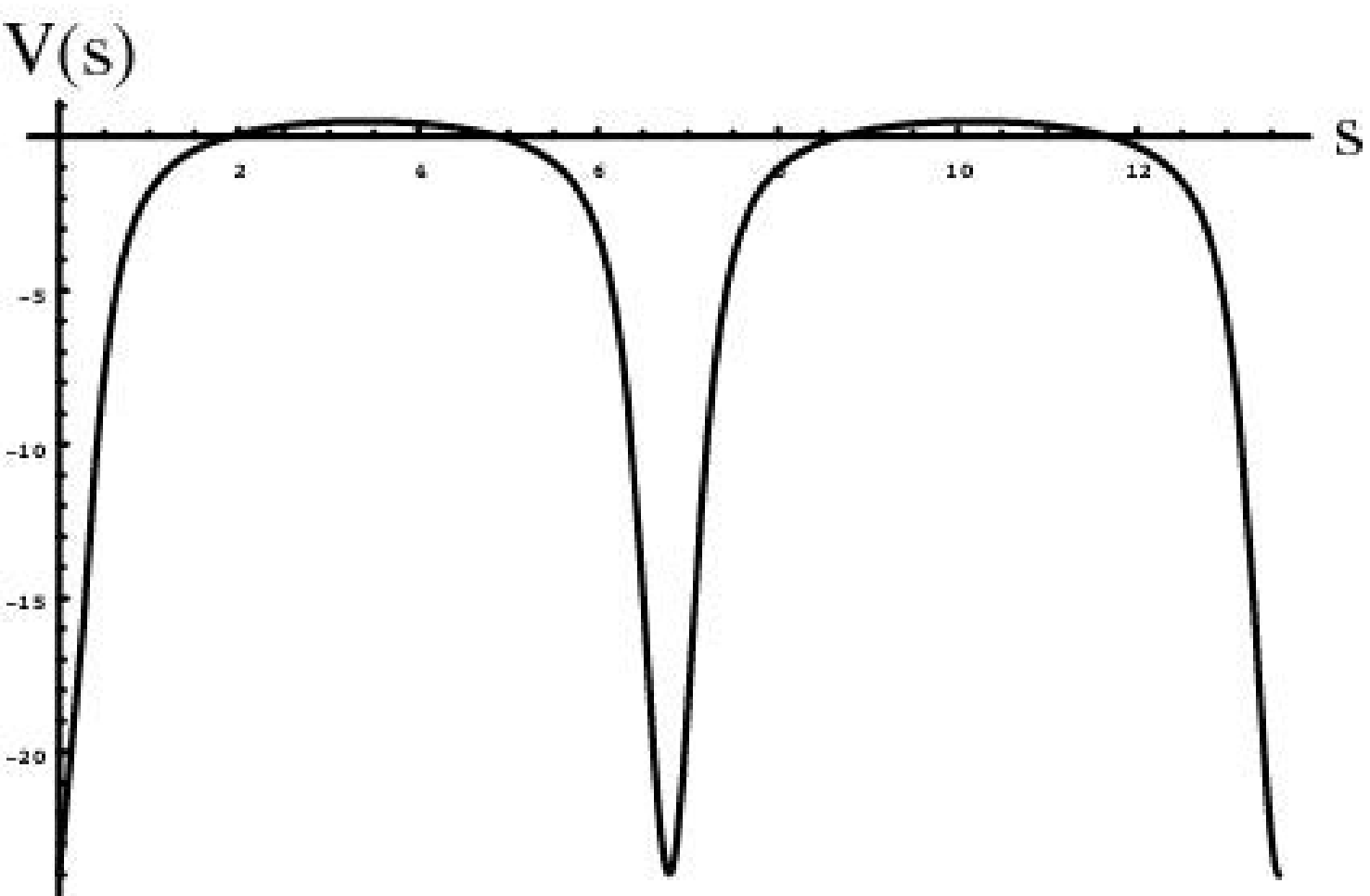,height=2in}}
\caption{The effective potential, $V(s)$ in the operator in
(\ref{schrod1})}
\label{fig:poten}
\end{figure}

\begin{figure}
\centerline{\epsfig{file=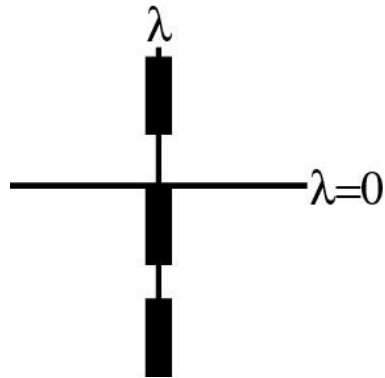,height=2in}}
\caption{The spectrum of the potential shown in Figure \ref{fig:poten}}
\label{fig:bands}
\end{figure}

When we are dealing with deep potential wells and large negative
eigenvalues, the usual numerical integration of a differential
equation over long intervals gives erroneous result.  The reason is
that as we integrate along a given path, the error starts growing
exponentially and the longer is the distance, the more unreliable the
final answer is. To avoid this, we only integrated numerically over
half of a loop and with the help of the transfer matrix, we calculated
the eigenfunctions in other regions. If we have a potential in the interval $ 0
\le x \le L$, and if the potential has reflection symmetry about
$x=L/2$, we can express a solution $\psi(x)$ and its derivative at
$x=L/2$ in terms of the two ``primary'' functions, $\Phi_{1}(x)$ and
$\Phi_{2}(x)$, and their derivatives, at $L/2$, as follows
\begin{eqnarray}
\left( \begin{array}{l} \psi(L/2) \\ \psi^{\prime}(L/2) \end{array}
\right)& =& \left(\begin{array}{ll} \Phi_{1}(L/2) &
\Phi^{\prime}_{1}(L/2) \\ \Phi_{2}(L/2) & \Phi_{2}^{\prime}(L/2)
\end{array} \right) \left( \begin{array}{l} \psi(0) \\
\psi^{\prime}(0) \end{array} \right) \nonumber \\ &\equiv &
\textbf{T} \left( \begin{array}{l} \psi(0) \\ \psi^{\prime}(0)
\end{array} \right)
\label{transfer1}
\end{eqnarray}
Here, the functions $\Phi_{1}(x)$ and $\Phi_{2}(x)$
satisfy the equation in the interval. They also satisfy the boundary
conditions
\begin{eqnarray}
\Phi_{1}(0) & = & 1
\label{phi10} \\
\Phi_{1}^{\prime}(0) & = & 0
\label{phi1prime0} \\
\Phi_{2}(0) & = & 0
\label{phi20} \\
\Phi_{2}^{\prime}(0) & = & 1
\label{phi2prime0}
\end{eqnarray}
To find the function at the end of the interval, $x=L$, we reverse
its sign in the middle and multiply by $\textbf{T}^{-1}$, That is
\begin{equation}
\left( \begin{array}{l} \psi(L) \\ \psi^{\prime}(L) \end{array}
\right) = \textbf{R}\textbf{T}^{-1}\textbf{R}\textbf{T}\left( \begin{array}{l}
\psi(0) \\ \psi^{\prime}(0) \end{array} \right)
\label{acrossL}
\end{equation}
where
\begin{equation}
\textbf{R} = \left( \begin{array}{rr} 1 & 0 \\ 0 & -1 \end{array}
\right)
\label{Rdef}
\end{equation}
Effectively, we have a potential of the form of Figure 
\ref{fig:poten} and an interval
equal to $2L$.

To get anywhere along two loops,  we use
the appropriate combination of the above transfer and slope-reversing
matrices.
It is useful to construct a look-up table of the ``fractional''
transfer matrix $\textbf{t}(x)$, where
\begin{equation}
\textbf{t}(x) = \left(\begin{array}{ll} \Phi_{1}(x) &
\Phi^{\prime}_{1}(x) \\ \Phi_{2}(x) & \Phi_{2}^{\prime}(x)
\end{array} \right)
\label{fract1}
\end{equation}
where $0 \le x \le L/2$.  With the use of this matrix, we can
construct solutions throughout the interval.  For instance, to obtain
the solution in the interval $L/2 \le x \le L$, one makes use of the
following relationship:
\begin{equation}
\left( \begin{array}{l} \psi(x) \\ \psi^{\prime}(x) \end{array}
\right) = \textbf{R}\textbf{t}(L-x)\textbf{R}\textbf{T}^{-1}
\textbf{R}\textbf{T}
\left( \begin{array}{l}
\psi(0) \\ \psi^{\prime}(0) \end{array} \right)
\label{alittlebit}
\end{equation}
Or, to obtain the solution in the interval $3L/2 \le x \le 2L$, one
utilizes:
\begin{equation}
\left( \begin{array}{l} \psi(x) \\ \psi^{\prime}(x) \end{array}
\right) = \textbf{R}\textbf{t}(2L-x)\textbf{R}\textbf{T}^{-1}
\textbf{R}\textbf{T}\textbf{R}\textbf{T}^{-1} \textbf{R}\textbf{T}
\left( \begin{array}{l}
\psi(0) \\ \psi^{\prime}(0) \end{array} \right)
\label{long}
\end{equation}
Equation (\ref{long}) may appear very complex.  However, it saves a
considerable amount of computational time and leads to a reliable
answer.  One good measure of accuracy of our answer is the Wronskian
of the two independent solutions of the differential equation, which
was found to be constant, as expected, along the interval.

\end{appendix}

\begin{figure}
\centerline{\epsfig{file=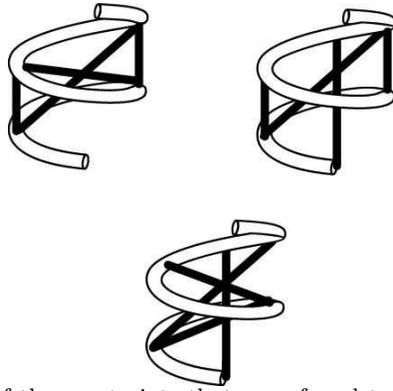,height=2in}}
\caption{A figurative depiction of some of the constraints that were
found to not stabilize the ``nucleosome'' configuration against
fluctuations.}
\label{fig:wrong}
\end{figure}

\end{document}